\documentclass[aps,prb,showkeys,preprint,showpacs,12pt]{revtex4-1}
\usepackage{amssymb}
\usepackage{hyperref}
\usepackage{graphics,graphicx}
\usepackage{dcolumn}
\usepackage{amsbsy}
\usepackage{amsmath}
\usepackage{epsfig}
\usepackage{color}
\usepackage{textcomp}

\begin{document}
 \title{Magnetodielectric effect in nickel nanosheet-Na-4 mica composites}
\author{Sreemanta Mitra$^{1,2}$}
\email[]{sreemanta85@gmail.com}
\author{Amrita Mandal$^{1,2}$}
\author{Anindya Datta $^{3}$}
\author{Sourish Banerjee$^{2}$}
\author{Dipankar Chakravorty$^{1,\dag}$}
\email[]{mlsdc@iacs.res.in}
\affiliation{
$^{1}$
 MLS Prof.of Physics' Unit,Indian Association for the Cultivation of Science, Kolkata-700032, India.\\ }
\affiliation{
$^{2}$
Department of Physics, University of Calcutta, Kolkata-700009, India.\\}

\affiliation{
$^{3}$
University School of Basic and Applied Science (USBAS),Guru Govind Singh Indraprastha University,New Delhi, India\\}

\begin{abstract}
 Nickel nanosheets of thickness 0.6 nm were grown within the nanochannels of Na-4 mica template. 
The specimens show magnetodielectric effect at room temperature with a change of dielectric constant as a
 function of magnetic field, the electric field frequency varying from 100 to 700 kHz. A decrease of 5{\%}
 in the value of dielectric constant was observed up to a field of 1.2 Tesla. This is explained by an inhomogeneous
 two-component composite model as theoretically proposed recently. The present approach will open up synthesis of various
 nanocomposites for sensor applications.
\end{abstract}
\maketitle

Multiferroic materials \cite{1,2} have been shown to exhibit a coupling between the dielectric property and the static magnetic field \cite{3,4,5,6}.
 Magnetodielectric effect has been studied for various materials e.g, manganese oxide \cite{5} double perovskites \cite{7}. There has been a number
 of evidence given in the literature showing that the magnetodielctric effect in suitable composites could arise due to the presence of 
inhomogeneities  in the system concerned \cite{8,9,10}. The present work has been motivated by a recent theoretical analysis, which predicted that
 such an effect could be possible in a two component two dimensional composite media, the components having different conductivities \cite{11}. 
Two-dimensional crystals of various materials have been the subject of investigation of late because of unusual
 properties expected in such systems \cite{12,13,14,15}. Chemical exfoliation of materials having layered structures has been adopted
 for generating flakes with few layers. A simple rubbing of fresh surfaces of suitably chosen crystals led to the formation 
of single layers of crystals \cite{15}. Recently two dimensional single crystalline nickel of triangular and hexagonal shape have been 
grown by a solution phase method \cite{16,17}. These have thicknesses around 6 nm, and edge length of the order of 15.4 nm.
\par
We have taken a template based approach to grow two dimensional silver \cite{18} and compounds like $BaTiO_{3}$ \cite{19} and GaN \cite{20,21}.
 For this purpose Na-4 mica, having the chemical composition $Na_{4}Mg_{6}Al_{4}Si_{4}F_{4}O_{20},xH_{2}O$, was used as the template. Na-4 mica has
 nanochannels within its crystal structure having a thickness of 0.6 nm \cite{22,23}.    
The structural characteristics of Na-4 mica allow the presence of four cations per unit cell. 
These can be exchanged with suitable ions which then contribute to the formation of the target compound after proper heat treatment.
 Because of a small thickness of the grown material, a much higher aspect ratio can be induced in it than that reported so far. 
In this letter, we report on the growth of Ni nanosheets within Na-4 mica channels. The composites exhibit magnetodielectric characteristics.
 The results are consistent with the inhomogeneous conductor model referred to earlier. The details are reported in this letter.  
Primarily Na-4 mica was synthesized by usual sol-gel technique. Synthesis of Na-4 mica has been described elsewhere \cite{19,24}.
  The Na-4 mica powder was kept inside a saturated solution of $Ni(NO_{3})_{2},6H_{2}O$  (as obtained from E.Merck (India) Ltd.) 
for two and a half months, at an elevated  temperature ~ 333K in an air oven, for the ion exchange reaction, $2Na^{+}\Leftrightarrow Ni^{2+}$ to take place.
 The mixture was stirred from time to time. Besides the ion exchange process, the nanochannels were filled with $Ni(NO_{3})_{2}$  molecules, which
 had gone into the channels by means of diffusion. The powder was then dried and washed thoroughly with deionised water several times,
 so that no $Ni(NO_{3})_{2}$ molecule was left on the surface of the powdered sample. After few repetitions no trace of nickel was present in the
 filtrate as found by chemical group test. Dimethyl Glyoxime was added to the filtrate and no red colored precipitate was found which signified
 that there was no nickel in the filtrate. Then the sample was subjected to a reduction treatment. The powder, in an alumina boat, was brought
 to 1273 K in nitrogen atmosphere inside a tubular furnace and then it was subjected to Hydrogen ($H_{2}$) gas flow at that temperature for 1 hour.
 Hydrogen reduced the $Ni^{2+}$ ions in the nanochannels to form nickel. The sample was brought back to room temperature by furnace cooling in nitrogen atmosphere.
\par
X-ray diffraction study on the samples was performed using monochromatic $Cu K_{\alpha}$ radiation (wavelength 0.1540 nm), (Bruker D8 XRD SWAX) starting from 2 $\theta$=10$^{o}$
 to 80$^{o}$. 
To study the microstructure, nickel nanosheets were extracted from the Na-4 mica channel by etching the composite sample with 
40{\%} HF aqueous solution and centrifuged in SORVALL RC 90 ultracentrifuge at 40,000 rpm for 30 minutes. The morphology and electron diffraction
 pattern were studied by transmission electron microscope (JEOL 2010) operated at 200kV.
For dielectric and magnetoelectric coupling measurements the composite powder was cold pressed by applying a pressure of $5tons-cm^{-2}$ to form a pellet
 of 1cm diameter and 1 mm thickness. Silver paint (supplied by M/S Acheson Colloiden, Netherland) was applied on both faces of the samples to form electrodes.
 The capacitance change as a function of applied magnetic field (using a large water cooled electromagnet) was measured using an Agilent E4980A precision LCR meter.
 In the latter measurement the orientation of the applied magnetic field was perpendicular to the electric field for determining the value of the dielectric constant.\\

From X-ray diffraction pattern of the composite sample, presence of both nickel and Na-4 mica can be confirmed with their corresponding lattice planes,
 marked on the figure 1 \cite{25} .
 Figure 2(a) shows the transmission electron micrograph of nickel nanosheets grown within Na-4 mica. Figure 2(b) gives the high resolution transmission electron micrograph
 of one of the nanosheets. The interplanar spacing obtained from the image is 0.20 nm, which is in agreement with the $d_{hkl}$ value for the (111) plane in nickel. The presence
 of nickel phase is further confirmed by figure 2(c), which shows the selected area electron diffraction pattern.  The $d_{hkl}$ values are calculated from figure 2(c) and
 summarized in table 1. These are compared with Joint Committee on Powder Diffraction Standards (JCPDS) data (file no.04-0850) for nickel and Na-4 mica \cite{25}. 
It is evident that the composite contains both the nickel and Na-4 mica phases confirming the growth of nickel films within the Na-4 mica structure. 
\begin{figure}
\centering
\includegraphics{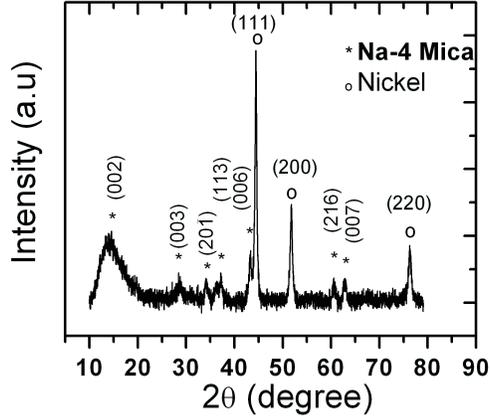}
\caption{ X-ray diffraction pattern of the composite sample comprising Na-4 mica and nickel.}
\label{fig.1}
   \end{figure}  
\begin{figure}
\centering
\includegraphics{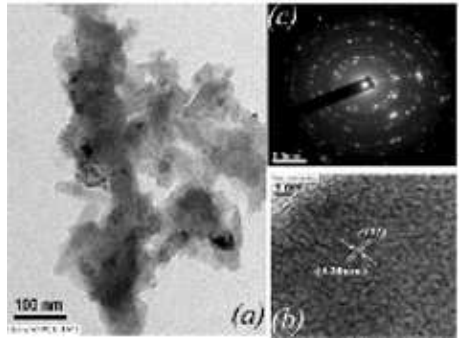}
\caption{(a) Random assembly of nickel nanosheets. (b) High resolution lattice image from a nickel nanosheet. (c) Selected area electron diffraction pattern of (a).}
\label{fig.2}
   \end{figure}
 \begin{table}
\caption{Interplanar spacing estimated from electron diffraction data and JCPDS file respectively.}
\label{tab.1}
\begin{center}
\begin{tabular}{lcr}
 \hline
Observed & Na-4 mica & Nickel\\ (nm) & (nm) & (nm)\\
 \hline
0.25 & 0.244 & $-$ \\
0.20 & $-$ & 0.203 \\
0.15 & 0.15 & $-$ \\
0.12 & $-$ & 0.124 \\
0.10 & $-$ & 0.101 \\
0.09 & $-$ & 0.088 \\
\hline
\end{tabular}
\end{center}
\end{table}
We have performed the dielectric permittivity measurement of the composite specimen and delineated 
the real and imaginary components of the same. Figures 3 and 4 show the frequency dependence of the 
real $(\epsilon ^{'})$ and imaginary  $(\epsilon ^{''})$ parts respectively, measured at different temperatures. It is seen that 
the real part  $(\epsilon ^{'})$ decreases from a high value as the frequency is increased, whereas the imaginary part  $(\epsilon ^{''})$ clearly shows
 a Debye like relaxation peak \cite{26} which shifts to higher frequencies as the temperature is increased. It is to be noted that
 the lines in figures 3 and 4 are drawn to guide the eye. The composite material shows dielectric dispersion and we have shown 
this more clearly in fig. 5 where a Cole-Cole diagram \cite{26} has been plotted. It is evident that the semi-circle characterizes the
 dielectric data. This can be explained on the basis of a heterogeneous dielectric having laminae with different values of 
dielectric permittivity and electrical conductivity. It was shown earlier that such a system exhibits a dielectric dispersion 
with a single relaxation time \cite{27,28}. The laminae in our case comprise the nickel films within the nanochannels of Na-4 mica
 and the mica blocks constituted by silicon-oxygen tetrahedra and aluminum-oxygen octahedra. Thus the dielectric measurement
 and the analysis thereof establish the fact that this composite is electrically inhomogeneous. 
\begin{figure}
\centering
\includegraphics{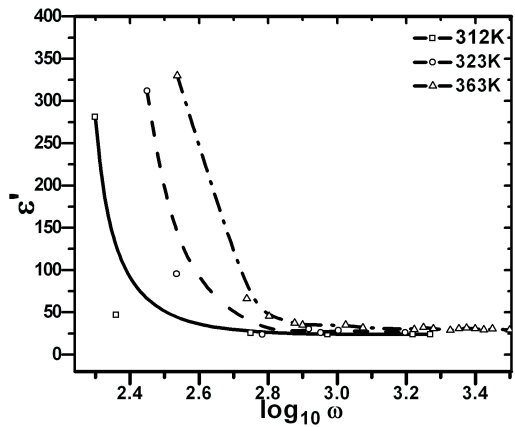}
\caption{ Variation of real part of permittivity  $(\epsilon ^{'})$  for the nanocomposite as a function of frequency measured at different temperatures.}
\label{fig.3}
   \end{figure}  
\begin{figure}
\centering
\includegraphics{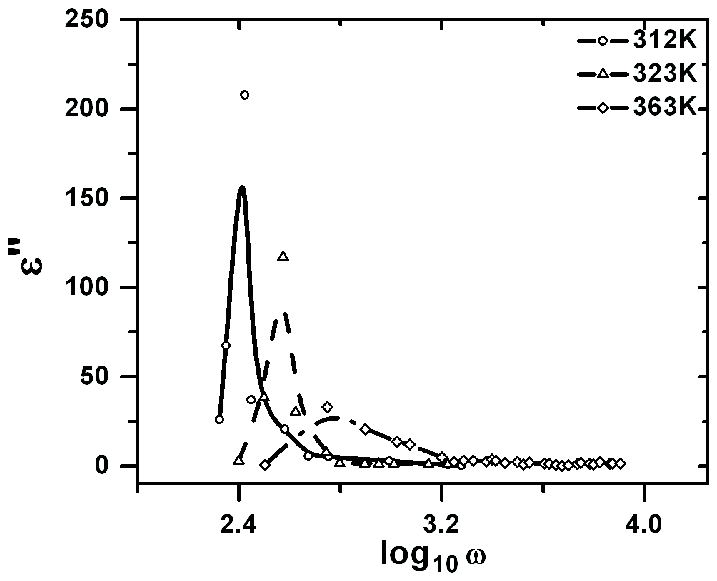}
\caption{ Variation of imaginary part of permittivity  $(\epsilon ^{''})$  for the nanocomposite as a function of frequency measured at different temperatures.}
\label{fig.4}
   \end{figure}  
\begin{figure}
\centering
\includegraphics{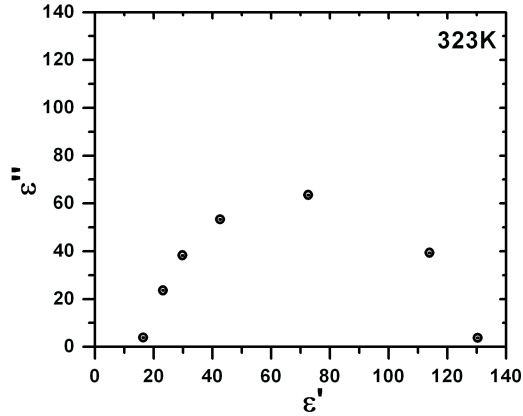}
\caption{Cole-Cole diagram for the nanocomposite measured at 323 K. }
\label{fig.5}
   \end{figure} 
We have investigated a possible magnetodielectric effect in our system. 
This was motivated by a recent theoretical work in which it was shown that magnetocapacitance effect 
can be observed in a two component composite medium [11]. The physical mechanism is essentially based on 
the accumulation of space charge layer at the boundary of the conducting and non conducting layers respectively.
 This causes an effective dipole moment within the system which is reflected in the corresponding dielectric constant
 of the material. Application of a magnetic field causes movement of these charges perpendicular to both the electric and
 magnetic field directions (Hall effect). This brings about a lowering of the space charge and hence a decrease in the value of
  the dielectric constant.  The dielectric permittivity of an inhomogeneous medium consisting of a purely resistive and purely capacitive
 region connected in series has been shown to be given by

\begin{equation}
 \epsilon_{c}({\omega})={\epsilon}\frac {(1+i{\omega}{\tau})} {\sqrt{i{\omega}{\tau}[(1+i{{\omega}{\tau})}^{2}-{({\omega}{\tau}{\beta})}^{2}]}}
\end{equation}

where, $\epsilon_{c}({\omega})$ is the effective dielectric permittivity, $\omega$ is the angular frequency of the applied electric field,
 $\epsilon$ the dielectric constant of the region in Na-4 mica containing the covalently bonded blocks comprising  Si, Al, Mg, and O, $\tau=\epsilon\rho$, $\rho$ 
  being the resistivity of the nickel nanofilm, and $\beta=\mu H$; $\mu$ being carrier  mobility and H the  applied magnetic field. Figure 6 
shows the variation of real part of effective dielectric permittivity $\epsilon_{c}({\omega})$ of our sample as a function of magnetic field H. It is seen
 that $\epsilon_{c}({\omega})$ decreases as H is increased. We have fitted the experimental data to equation (1) using $\mu$ as the parameter. A value of 
      $\rho = 6.14(10^{-6} Ω-cm)$ \cite{29} for bulk nickel and $\epsilon$ = 17 \cite{24} for Na-4 mica were used in these calculations. 
The theoretically fitted points are also shown in figure 6. It is evident that the dielectric constant of the composite
 system shows an increase with the lowering of frequency. This is a characteristic feature of Maxwell-Wagner polarization effect \cite{26}.
 Our dielectric data substantiate this aspect of the nanocomposite we have studied here.
\begin{figure}
\centering
\includegraphics{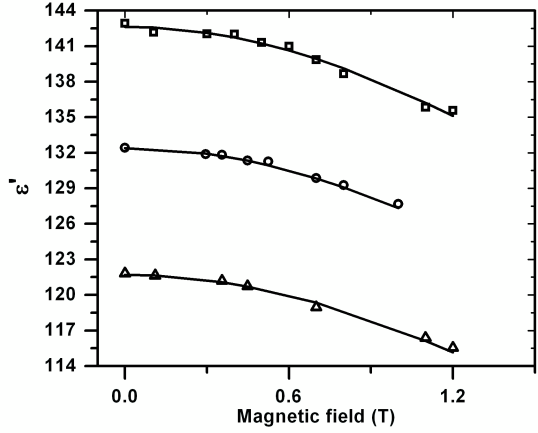}
\caption{Variation of dielectric constant of the nanocomposite with magnetic field at different electric field frequencies}
\label{fig.6}
   \end{figure} 
\begin{figure}
\centering
\includegraphics{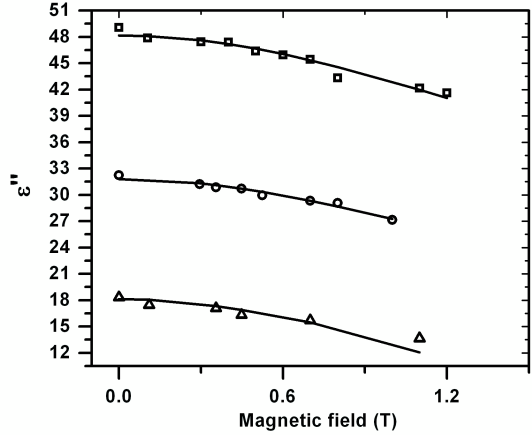}
\caption{Variation of dielectric constant of the nanocomposite with magnetic field at different electric field frequencies}
\label{fig.7}
   \end{figure} 

 In figure 7 is shown  the experimental and theoretically fitted data for the imaginary part of dielectric permittivity as a function 
of applied magnetic field. It is to be noted that the percentage changes in the real and imaginary parts of dielectric permittivity 
were found to be in the ranges of $ 3.6 $ to $ 5.2 {\%} $ and $ 15.2 $ to $ 25.4{\%}$ respectively.    
 A satisfactory agreement is observed between the two sets of values for all frequencies at which measurements were carried out. 
The theoretical analysis and successful fitting of the experimental data to the model of a two component two-dimensional composite medium validates
 the space charge polarization effect we have invoked in our discussion. 
The extracted value of $\mu$ is found to be 0.28 $m^{2}volt^{-1}sec^{-1}$.   
The dimension of $\mu$ should be $(Tesla)^{-1}$. The latter which is $(Weber)^{-1}(m^{2})$ can be related by Faraday’s law of electromagnetic induction to $m^{2}volt^{-1}sec^{-1}$.
 This is much larger than the mobility of carriers in the bulk nickel viz, $0.56(10^{-3} m^{2}volt^{-1}sec^{-1})$. We argue that in our case the conducting phase
 i.e. nickel has got higher mobility due to its ultrathin configuration. Such an increase in mobility has been shown to occur in ultra-thin films \cite{30}.
 This is related to a decrease in carrier density and a consequential increase in the average time between carrier collisions under the influence of an applied 
electric field. As the number of collisions decreases the movement of the carriers becomes much easier. This is reflected in an increase in mobility. 
It may be mentioned that the magnetodielectric effect studied here was measured with an electric field of frequency in the range 100 kHz to 700 kHz at room temperature
 with a maximum magnetic field applied as 1 Tesla. These are conditions ideal for the construction of a magnetic sensor as pointed out by earlier authors \cite{8}. 
\par
Lastly, we need to point out that the present nanocomposite system contains nickel and as such the latter could be expected to contribute independently
 to its magnetodielectric behaviour. Our measurements show this material to be ferromagnetic, exhibiting superparamagnetic behavior above 428 K. 
Details of these properties will be reported elsewhere. It has not been possible to directly measure the magnetoresistance of the nickel films present in our system.
 However, we have used  the magnetoresistance data of 80nm thick nickel film as reported in literature \cite{31}. Using equation (1) and incorporating a negative change
 in resistivity of 0.3{\%} at an applied magnetic field 1.2 T we estimated a change in dielectric constant of 0.14{\%}. This is an order of magnitude less
 than the changes observed in our studies. Hence the effect of magnetoresistance of nickel on the magnetodielectric properties of our nanocomposite sample is negligible.
\par
 In summary, nanosheets of nickel with thickness equal to 0.6 nm have been grown within the nanochannels of Na-4 mica. 
The composites show dielectric relaxation behaviour expected of a two-phase laminar conductor. The specimens exhibit 
a magnetodielectric effect with a change of dielectric constant as a function of magnetic field of $5{\%}$ up to a field of 1.2 Tesla.
 The measurements were carried out at room temperature and with an electric field of frequency in the range 100 to 700 kHz. 
The results are explained on the basis of an inhomogeneous two-component composite model. This strategy of specimen preparation will
 lead to development of magnetic sensors for various applications.
\bibliography{Bibliography}

\end{document}